%% file: paper.tex
\documentclass{llncs}
 
\author{Constantin Jucovschi}
\title{Cost-Effective Integration of MKM Semantic Services into Editing Environments}

\usepackage[hyperref=auto,style=alphabetic,maxnames=3]{biblatex}

\bibliography{kwarcpubs,kwarc,kwarccrossrefs,extpubs,extcrossrefs}

\def\march{ReDSyS}

\usepackage[hide]{ed}
\usepackage{paralist}
\usepackage{stex-logo}
\usepackage{listings}
\usepackage{graphicx}
\usepackage{amssymb}

\lstMakeShortInline[basicstyle=\tt]|
\usepackage{lstomdoc}

\begin{document}
\institute{Jacobs University Bremen, Germany}

\maketitle

\input{abstract} 
\input{intro} 
\input{related} 
\input{architecture} 
\input{implementation} 
\input{conclusion}

\printbibliography

\end{document}

%% file: abstract.tex
\begin{abstract}
  Integration of MKM services into editors has been of big interest in
  both formal as well as informal areas of MKM. Until now, most of the
  efforts to integrate MKM services into editing environments are done
  on an individual basis which results in high creation and
  maintenance costs.

  In this paper, I propose an architecture which allows editing
  environments and MKM services to be integrated in a more efficient
  way. This is accomplished by integrating editors and services only
  once with a real-time document synchronization and service
  broker. Doing so simplifies the development of services, as well as
  editor integrations. Integrating new services into an arbitrary
  number of already integrated editors can then take as little as 3-4 hours
  of work.
\end{abstract}
%%% Local Variables: 
%%% mode: latex
%%% TeX-master: "paper"
%%% End: 

%% file: intro.tex
\section{Introduction}
\label{intro}

Integration of MKM services into editors has been
of big interest in both formal as well as informal areas of MKM.
In the formal area of MKM, it becomes more and more important to 
have an efficient way to work with formal documents and 
pass information between interactive
provers and the user. Examples of useful services are: showing type 
information when hovering 
over an expression, navigating to definitions of symbols, and 
supporting editor features like folding inside long formulae.

Informal MKM needs editing support to make it easier for 
authors to create 
semantically annotated documents. This can mean integration
of Natural Language Processing (NLP) services to e.g. spot mathematical 
terms as well as hiding (or folding) 
existing semantic annotations in order to provide a better reading experience.

Until now, most of the efforts to integrate MKM services into 
editing environments were done on an individual basis, namely, some
service $X$ was integrated into an editing environment $Y$.
Obviously it would have been more efficient to integrate service
$X$ into all editing environments $Y$ where this service
makes sense. The problem is that developing and maintaining 
such integrations requires a lot of effort and hence MKM services 
are usually integrated with maybe one or two editing environments. 

The high cost of semantic service integration has an especially 
negative impact on informal MKM because its users, the authors 
of mathematical texts, write mathematics in a handful of different programs
and operating systems. Even if one chooses to support the 5 most used
editing environments, it would still be too expensive to maintain 
10 semantic service integrations. Clearly, a different integration strategy
is called for.

In this paper, I propose an architecture allowing tight integration
of authoring services into the authoring process. It enables distributed 
services, running on different platforms and hardware, to be
notified of changes made by the user to a certain document. Services can 
then react to these changes by modifying/coloring document's text as well as 
enriching parts of the text with extra (invisible to the user) semantic
annotations. This architecture can accommodate both 
\begin{inparaenum}
\item reactive services like syntax highlighting giving user the
  illusion that the service runs natively in the editor as well as,
\item time-consuming services like Natural Language Processing tasks,
  without requiring the user to wait while text is processed. 
\end{inparaenum}

In the next section I discuss in more detail what type of services would
fit the presented framework, and compare it to other existing MKM frameworks.
In section \ref{architecture}, I present the proposed
integration architecture. To validate
the architecture I implemented four services and integrated them
in two editors. In section \ref{impl}, I briefly describe these 
services, and give some high level technical details on how new
services can be built. Section \ref{concl} concludes the paper.

%%% Local Variables: 
%%% mode: latex
%%% TeX-master: "paper"
%%% End: 

%% file: related.tex
\section{Aims and Scope of Integration}
\label{related}
The task of tool integration is a very complex and multi-faceted one. 
Many frameworks and technologies \cite{Wicks2004} have been proposed to
integrate tools, each optimizing some aspects of tool integration
and yet, none of them is widely adopted. The current paper does not 
attempt to create a framework to integrate all possible editors 
with all possible services. It considers only pure text editors and integrates only
services that participate in the editing/authoring process. The scope of 
the framework is purposefully kept relatively small so that it solves 
a well-defined part of editor-service integration problem and can 
eventually be used along with other integration frameworks. 

In the next section, I would like to make more explicit the types of
services that are included in the scope of the framework.
In section \ref{levels_of_integration} I analyze what integration
strategies will be used to achieve optimal integration. This will later help 
me in section \ref{comparison} to differentiate the framework more clearly from 
other existing frameworks. 

\subsection{Targeted Authoring Services}
\label{targeted_services}
In this paper by ``authoring service for text-based documents'' 
(abbreviated as ``authoring service'' or just ``service''), I mean
a service which provides some added value to the process of authoring 
the text document by:
\begin{compactenum}
\item reacting to document changes and giving feedback e.g. by
  coloring/underlining parts of text,
\item performing changes to the document e.g. as result of an explicit
  request to autocomplete, use a template, or fold some part of
  text/formula.
\end{compactenum}

The strategy to achieve editor-service integration is to require authoring services 
to be agnostic of advanced editor features like the ability to fold 
lines or embed images/MathML formulae. Services should only be allowed to 
assume a relatively simple document model that allows a 
limited set of operations (both described in section \ref{model_changesets})
and which are supported by most editing environments. Likewise,
services should only assume a limited set of possible interactions with 
the user (see section \ref{uimodel}). 

A way to decide whether a certain service is in the scope 
of the presented integration framework is to analyze whether it can
be realized conforming to the limited document model and interaction 
possibilities. \ednote{authoring services that cannot be implemented 
with the framework?}

\subsection{Levels of Integration}
\label{levels_of_integration}
One of the classical ways used to describe and compare integrations
was proposed by Wasserman \cite{Wasserman1990} and later improved by 
Thomas and Nejmeh \cite{Thomas1992}. They proposes 4 dimensions along 
which an integration can be analyzed, namely: presentation, 
data, control and process integration dimensions. Presentation integration 
accounts for the level at which tools share a common ``look and feel'', 
mental models of interaction, and interaction paradigms. Data integration 
dimension analyzes how data is produced, shared and kept consistent
among tools. Control integration dimension analyzes the level to which 
tools use each other's services. The process integration dimension describes
how well tools are aware of constraints, events and workflows taking place
in the system. 

Note that high or low level of integration does not 
reflect the quality of some integration. Low integration
level in some dimension only suggests that those tools can be
easily decoupled and interchanged. High integration level, on the other
hand, suggests that tools connect in a deeper way and can enable
features not possible otherwise. Experience suggests 
that, low level of integration require less maintenance 
costs in the long run and should be used whenever possible. 

According to description of the targeted authoring services, I derived 
the integration levels that need to be supported in each integration dimension. Namely we need:
\begin{compactenum}
\item medium-high presentation integration level, because services
  need to be able to change the text/colors of the document as well as 
  interact with the user using some high level interaction paradigms.
  While the type of changes/interactions a service can perform are limited,
  a service should be able to perform these actions unrestricted by other
  components. 
\item high data integration level due to the fact all service are 
  distributed and still need to be able to perform changes to the edited document. 
  Hence synchronization and data consistency mechanisms are needed.
\item low control integration as the framework should only give support for
  integrating services with editors. Service-service integrations
  are outside the scope of integration.
\item low process integration mainly due to the fact that services
  are expected to be mostly stand-alone and the workflows should only
  involve an editor and a service. 
\end{compactenum}

\subsection{Comparison to other MKM Integrations}
\label{comparison}
A lot of integrations combining several tools have been 
developed in MKM. In the context of the current paper only several
types of integrations are interesting:
\begin{compactenum}
\item frameworks that enable integration of services into authoring
  process in a consistent, service independent way,
\item integrations which allow multiple services to listen/react to
  document changes.
\end{compactenum}

The Proof General (PG) framework along with the PGIP message 
protocol \cite{AspinallLW07} constitute the base for several popular 
integrations between editors (e.g. emacs, Eclipse) and interactive 
provers like Isabelle, Coq and HOL. PG framework differs from presented
framework in the following ways:
\begin{compactenum}
\item low presentation integration --- consisting of changing the
  color of text regions according to prover state, accompanied by
  locks of regions under processing. The later is not a typical
  interaction a user would expect. Integrated provers cannot directly
  influence the display or interact with the users. Hence only the
  broker component is directly integrated from a presentation
  integration point of view.
\item low data integration --- editors and services communicate mainly
  by sending parts of the source text to the prover as result of the
  user changing regions of text. This data never gets changed by
  services and hence no synchronization or consistency checking is
  needed.
\item medium control integration --- the broker uses different
  protocols to talk with displays and provers. Additionally, it has the
  role of orchestrating interaction between them.
\end{compactenum}

The effort of Aspinall et al. \cite{Aspinall2006} further extends PG 
architecture to integrate rendering processors 
(e.g. \LaTeX{}) and possibly other tools (e.g. code generators). 
These newly integrated components are loosely integrated by running
them on a (hidden from the user) updated version of the original 
document. The broker-prover integration becomes tighter due to the
documentation and script backflow mechanisms. Differences to the 
current framework are:
\begin{compactenum}
\item from the presentation integration perspective it extends the PG
  framework by two (or more) additional views of the document. These
  views are presented and updated in separate windows and provide the
  user with more focused views on the authored content.  These new
  views do not seem to provide additional interactions to the user. In
  conclusion, the presentation integration is tighter compared to PG
  but still relatively low.
\item data integration becomes tighter between the broker and prover
  components as a result of the backflow mechanisms but is still
  relatively low. The additional complexity due to the backflow is
  mostly in the broker component which needs to know where in the
  central document to integrate data coming from the prover. Data
  passed arround still never gets modified and hence no advanced
  synchronization or consistency checking is needed.
\item control integration is similar to the PG framework.
\end{compactenum}

The integration of provers and editors is done quite differently in 
PIDE~\cite{Wenzel:2010}. Instead of fixing a protocol encapsulating 
all the features a prover can provide (like PGIP does), a document
model is specified and the protocol to interact with that document model
is fixed. This has the advantage that editors need to know
much less about the provers features, and only need to provide
them with changes the user made to the document. Conceptually this is 
very similar to the approach taken in the current work. The difference
is that in the case of PIDE, the document model is shared only between two
entities (the editor and the prover) and that these entities share the
same running environment. The current architecture allows
several authoring services to listen and change the shared document and 
allows them to be distributed. 

%%% Local Variables: 
%%% mode: latex
%%% TeX-master: "paper"
%%% End: 
 

%% file: architecture.tex
\section{Editor Service Integration Architecture}
\label{architecture}
Creating and maintaining integrations between software programs 
is generally an expensive task. However, there are some well known
best practices for these tasks which have been proven to help a lot in reducing both creation
as well as maintenance costs. A good example is the integration
between database systems and hundreds of languages and frameworks. 
Some of the key aspects that make such integrations possible are:
\begin{compactenum}
\item[{\bf P1.}] {\bf Client-server architecture} allows
  clients and servers to be developed and executed on arbitrarily 
  different environments. 
\item [{\bf P2.}]{\bf Stable communication protocol} on the server side
  reduces maintenance costs and makes documentation more stable
  and complete.
\item [{\bf P3.}]{\bf Declarative API} is usually more stable as it requires 
  definition of a small set of primitives and some way to compose them. 
\end{compactenum}

\begin{figure}
  \centering
  The ReDSyS Architecture
  \includegraphics[width=12cm]{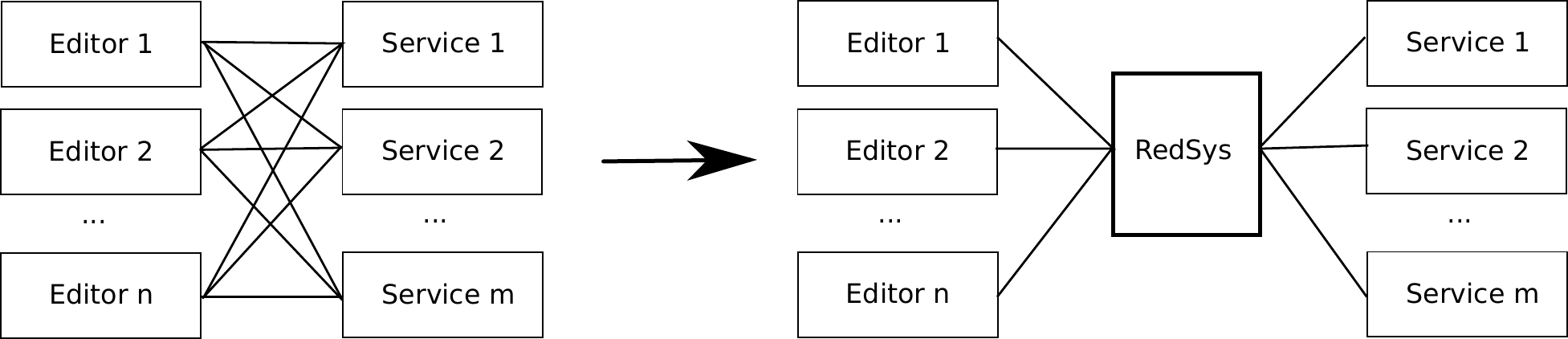}
  $n \cdot m$ integrations \hspace{1cm} vs \hspace{2.5cm} $n+m$ integrations
  \caption{Direct integrations between editors and environments
  vs indirect integration through \march{} component.}
  \label{fig:archchange}
\end{figure}

The goal of the current architecture is to integrate $m$ services 
into $n$ editors. Direct integration between editors 
and services (Figure \ref{fig:archchange}a) would 
require $n\cdot m$ integrations to be implemented. 
To reduce this number, I propose to create an independent Real-Time Document Synchronization and Service Broker (\march{}) component, 
as shown in Figure \ref{fig:archchange}, which complies with the practices {\bf P1}-{\bf P3},
and which integrates with each 
service and editor exactly once in a client-server manner (\march{} being
the server). In this way we only need $n$+$m$ integrations. 
The \march{} server API is expected to be stable 
(requirement P2) hence any upgrades of editors/services may require
adjustments in the integration only on the editor/service part. 

Section \ref{interaction_model} describes in more detail the 
\march{} component and how it integrates with editors 
and services. Section \ref{model_changesets} presents the shared
document model and shows how editing changes are represented. In section \ref{time_cons},
I discuss management of change issues that can appear when integrating time consuming 
services as well as make explicit some requirements for reactive services.
  Section \ref{uimodel}
presents the interaction model between users, editors and services.

\subsection{The Real-Time Document Synchronization and Service Broker}
\label{interaction_model}
The \march{} component has two main responsibilities:
\begin{compactenum}
\item provide a way for editors to trigger events to all or some
  subset of services,
\item allow editors and services to independently add/remove meta-data
  or text to the shared document in real-time.
\end{compactenum}

The first responsibility is typical for broker components. 
In our case, it makes it possible for editors to request 
autocompletion suggestions (from all services) or request 
the type of a symbol (from e.g. Twelf services for LF documents \cite{Pfenning91}). 

The second responsibility enables services to run 
independently, distributed on different systems and parallel to the editing process. So while the user is 
typing, a service might decide to start processing and integrate
the results back into the document when finished. Whenever a user 
changes the document, services get notified by the \march{} component and each
service has the freedom to decide whether to interrupt current 
processing (if available) or not. Thanks to real-time document editing solutions,
it is quite often possible to automatically merge service results
computed on older versions of the document into the current version.

\begin{figure}[h]
  \centering
  \includegraphics[width=12cm]{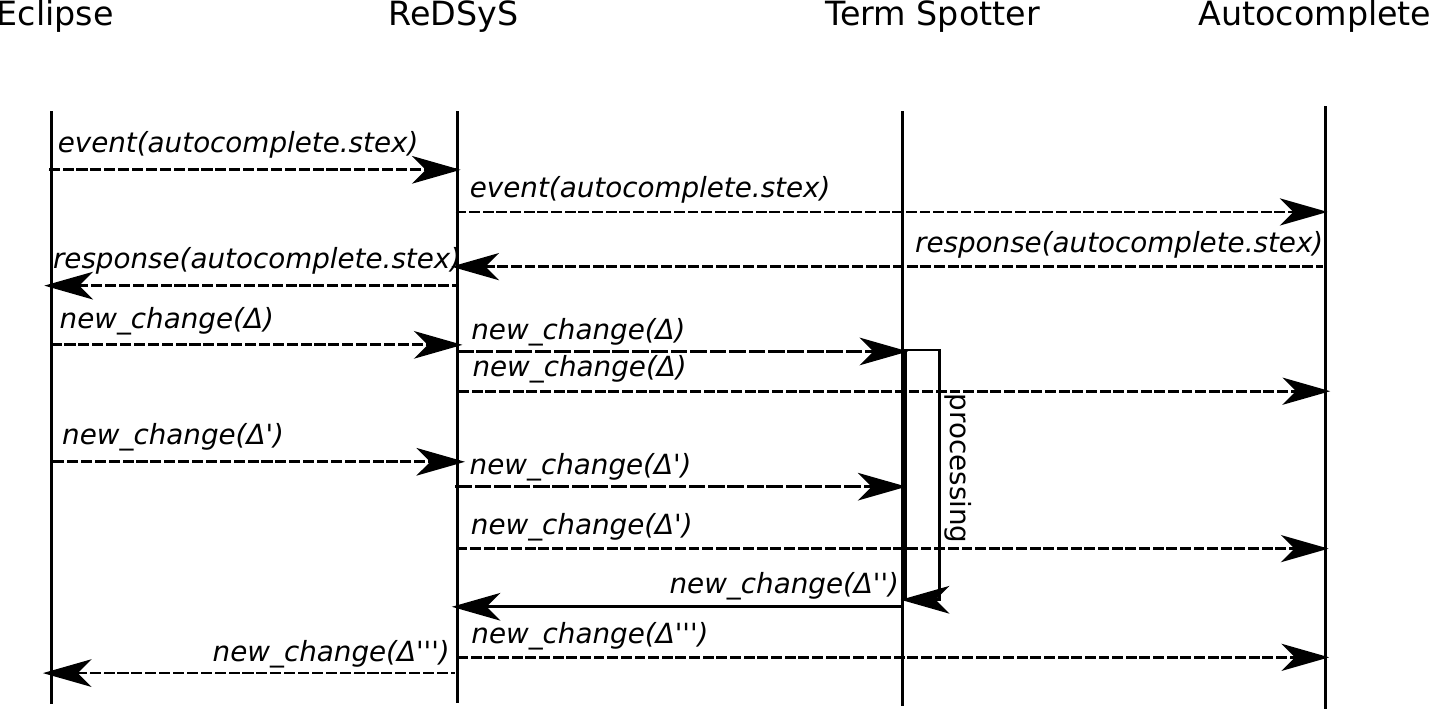}
  \caption{Communication diagram among architecture components after an autocomplete requests followed by term spotting.}
  \label{fig:comm_diag}
\end{figure}

To understand the interaction between editors, services and \march{} 
better, let me describe how \stex{}\cite{Kohlhase04:stex} editing in Eclipse\cite{Eclipse:web} can be integrated with 
a term spotting and an autocompletion service. You can follow the 
communication between components in Figure \ref{fig:comm_diag}.

The first step is to install the ``\stex{}-padconnector'' plugin into Eclipse which,
upon opening an \stex{} file, uploads it to the \march{} component 
and opens it in a typical Eclipse editing window. The opened document 
is, in fact, the shared document.  The \march{} architecture takes 
care of starting the semantic services.

The ``\stex{}-padconnector'' plugin is programmed so that, when the
user presses ctrl+space, it synchronously notifies (i.e. waits for the result) 
the \march{} architecture that an event with a predefined URI ``autocomplete.stex'' took place
and passes the current cursor coordinates as parameters. The autocomplete
service catches the event and based on the parameters decides on autocompletion
suggestions. The editor receives autocomplete suggestions from \march{} and displays them. 

Whenever the user changes the document, a changeset (or diff) is computed and
sent to \march{}. This passes on the changeset to all the other services so that they
have an up-to-date version of the document. Let us suppose that the Term Spotter 
service decides to start a relatively complex NLP processing task for of the new 
version (e.g. version 40) of the document. While processing, some other changeset comes 
to the Term Spotter service but it decides not to cancel the NLP task. The 
shared document has now version 41 but the NLP task has computed a changeset ($\Delta''$) 
which highlights new found terms based on document version 40. The Term Spotter service
sends change $\Delta''$ to \march{} also including information about the document
version on which the changeset is based on (i.e. 40). The \march{} component tries to
merge the changes and if it succeeds, sends a merged changeset to all the other components. 

\input{preliminaries}

\subsection{Time Consuming vs Reactive Services}
\label{time_cons}
The \march{} architecture can support both time consuming 
services (e.g. NLP tasks) that should not hinder the user 
from further editing of the document, as well as services 
that need to give user the impression that the service is 
running natively inside the editor (e.g. syntax highlighting). 

As described in the communication workflow in the previous 
section, time consuming services can start processing at any 
given point in time (e.g. version 40) and integrate their 
results automatically (e.g. at version 50) by creating 
and sending a changeset (call it X) based on the version 
when processing started (i.e. 40). The \march{} component 
has some merging strategies to integrate such changes 
but they will fail if the changes done in between i.e. 
from version 41-50 overlap with areas in changeset X.

This default behavior can be improved in a number of ways.
First, a time consuming service is still notified of 
changes done to the document even while it is processing.
Hence, it can interrupt the processing or restart it 
if the document was changed in the area currently under
processing. In this way, computing power to finish the 
processing (and then realize that it cannot be integrated) 
is not lost. The second solution is to try 
to incorporate incoming changes while or at the end of 
processing and ultimately create a changeset based on the 
latest version of the document. Hence a time consuming
service is responsible for it's own management of change.

To give users the feeling that reactive services run
natively in the editor, they need to be very optimized 
both in speed and in the size of the changesets 
they produce. These services might need to run at a rate
of 20 times a second in order to accommodate several
users editing in the same time. Hence it is very important 
that reactive services can cache results and start 
processing a document without reading all of it. Also 
the changesets that reactive services produce should be small 
and only change areas of the document that really need 
changing. For example, a bad syntax highlighting service
that creates a changeset recoloring the whole document
(and not only the parts the need recoloring) could invalidate
the processing of all the time consuming services.

\subsection{User Interaction Model}
\label{uimodel}
In the current framework, services and editors no longer 
integrate directly but they still need to interact, e.g. 
services might need to ask the user to disambiguate a 
mathematical term. Such interactions must be standardized 
so that all editors ask services to perform a certain 
action (e.g. autocomplete) in the same way. 

Every type of interaction between users, editors and 
services has a predefined URI. An example of such URIs 
is ``autocomplete.stex'' which, when broadcasted by
the \march{} component to the services, expects them 
to return \stex{} based autocompletion suggestions 
and then displayed to the user. Another example is
``contextmenu.spotter\_plugin.10''. This URI can be 
used inside an attribute (``ui'',''contextmenu.spotter\_plugin.10'') 
which, just like the (``bold'',''true'') attribute, 
can be applied to some part of the text in the shared document model 
presented in section \ref{model_changesets}. 

When the editor sees attributes having key ``ui'' and value 
prefixed with ``contextmenu.'', it knows to display a context menu 
when the text having this attribute is right-clicked. 
The menu items in the context menu are fetched from 
the ``spotter\_plugin'' component (i.e. the Term Spotter) 
and ``10'' is passed as an argument to identify which
context menu should be displayed. 

It is the editor that is responsible to understand interaction URIs and act accordingly.
That is why it is important to define interactions in a general and reusable manner so that
many services can take advantage of it. Currently, the set of predefined interaction URIs 
is relatively small and fits the use cases presented in the implementation section. 
However, it certainly needs to be revised and extended to fit a more general range 
of interactions. 

%%% Local Variables: 
%%% mode: latex
%%% TeX-master: "paper"
%%% End: 

%% file: preliminaries.tex
\subsection{Document Model and Changesets}
\label{model_changesets}
In this section I introduce the shared document model as it is important to 
understand what kind of information is shared among
services and how. This document model is the same as that of the Etherpad-lite system
\cite{etherpad-proto:web}.

We define a finite alphabet $A=\{\alpha_1,\alpha_2,...,\alpha_m\}$. A string $s_A=c_1c_2...c_n$ is 
a finite sequence of characters from alphabet $A$ and length $len(s_A)=n$ is defined
as the number of characters in that sequence. Let $Str_A$ be the set of strings over alphabet $A$.
For the sake of simplicity we will omit alphabet $A$ in notations when it can be 
unambiguously inferred from the context.

\def\AP{AP}
\def\atm{attmap}
\def\at{att}

An attribute pool $\AP=\left\{(id,(key, val))\in \mathbb{N}\times(Str \times Str) \mid id \textnormal{ - is unique}\right\}$ is a set of 
key-value pairs which also have a unique id assigned to each of them. This defines a function
$\atm:\mathbb{N}\to(Str \times Str)$ returning the key-value pair associated to a certain id.

A document $d=(text, \atm, \at)$ where $text\in Str$ represents the text in that document,  $\atm \in \mathbb{N}\to(Str \times Str)$
is the function associated to an attribute pool and $\at:\mathbb{N}\to Set(\mathbb{N})$ with $\at(i)=S$, $\|S\|<\infty$ 
specifies which set of attributes are assigned to character $i, 0\le i<len(t)$ in document's text.   

A change operation $d=(op, S, len, t)\in \{+, -, =\} \times Set(\mathbb{N}) \times \mathbb{N} \times Str$  
either
\begin{enumerate}
\item inserts (op=''+'') text $t$ of length $len$ and applies attributes in $S$ to each of the 
  inserted characters, or
\item deletes (op=''-'') $len$ characters (and their attributes) from a text ($t=""$ and $S=\emptyset$), or
\item leaves unchanged (op=''='') $len$ characters but applies attributes in $S$ (if $S\neq \emptyset$) to each of them.
\end{enumerate}

Editing changes inside a document are represented using lists of 
change operations $o=\{op_1op_2...op_k \}$ where $op_j$ are change operations 
such that no two consecutive operations have the same type and attribute sets (otherwise we can join them together). 

A changeset is defined as $c=\left(l, l', \atm, o \right)$, where $l$ is the length of the 
text before the change, $l'$ is the length of the text after the change, $\atm$ contains the
updated attribute pool (only new elements allowed) and a sequence of change operations to be 
applied on the text. 

To provide a better intuition for these notions, consider the document

\begin{equation}
  d=\left(\textnormal{"Math {\bf is} great"}, 
  \left\{ \begin{array}{l}
      0\to("bold","true") \\
      1\to("author","p1") \\
      2\to("author","p2")
  \end{array} \right\},
  \left\{ \begin{array}{l}
      0\leq i\leq4, \{1\} \\
      5\leq i <7, \{0,2\} \\
      7\leq i \leq 12, \{2\}
  \end{array} \right\}\right)
\end{equation}\\
which says that word ``Math'' was authored by ``p1'', the rest of the text by ``p2'' and the word ``is'' is bold.
Now consider a changeset 

\begin{equation}
  c=\left(13,12, 
  \left\{ \begin{array}{l}
      0\to("bold","true") \\
      1\to("author","p1") \\
      2\to("author","p2") \\
      3\to("author","") 
  \end{array} \right\},
  \left\{ \begin{array}{l}
      ("=", \{3\}, 1, ""),\\
      ("-", \emptyset, 3, ""),\\
      ("+", \emptyset, 2, "KM"),\\
      ("=", \emptyset, 9, ""),\\
  \end{array} \right\}\right)
\end{equation}\\
which when applied to $d$, would change the text to ``MKM {\bf is} great'', the word``MKM'' 
would have no author and the rest remains unchanged.
 
%%% Local Variables: 
%%% mode: latex
%%% TeX-master: "paper"
%%% End: 

%% file: implementation.tex
\section{Architecture Implementation}
\label{impl}

To validate the presented architecture and to prove its 
applicability to a wide range of semantic services and editors, 
I chose to integrate four \stex{} semantic services into two editors.
The semantic services were picked to address different integration
issues and are described in more detail in Section \ref{scenario}. 
Section \ref{api} gives a glimpse into the APIs that need to be
used in order to create a service. 
Finally, section \ref{extensibility} discusses extensibility and reuse issues
of the architecture. 

The editors I used to validate the architecture are: Eclipse (desktop based editor) and 
Etherpad's Web Client (web-based editor). 
To support the Eclipse based editor I implemented my own 
synchronization library with the \march{} component called 
jeasysync2. The real-time document sharing platform (\march{}) 
used in my implementation is an extension of the Etherpad-lite system and can be found at 
\url{https://github.com/jucovschi/etherpad-lite/tree/mkm}.

\input{scenarios}

\subsection{Libraries and APIs}
\label{api}
Currently one can create new services for my architecture using either
JavaScript or Java programming languages. JavaScript services are implemented
in Etherpad-lite's native plugin system. Java services should use the jeasysync2 
library. In both cases services must implement the following interface:
\begin{lstlisting}[language=java]
  void init(Changeset initialText, AttributePool pool);
  void update(Changeset lastChangeset, AttributePool newPool, ChangesetAcceptor csAcceptor);
\end{lstlisting}
The init method is called when initializing the service. The first parameter
is the changeset which, if applied to an empty text, generates the 
current document (note that attributes are included as well). 

The update method notifies the service of new updates. This is where the service
should decide whether to start/restart processing. The update
function is called asynchronously in separate threads so special care 
should be taken not to run into race conditions. The ChangesetAcceptor callback allows
the service to send changesets back to the \march{} component when processing is finished.

Creating changesets is easily done through a utility class called ChangesetBuilder with the methods:
\begin{lstlisting}[language=java]
  void keep(int noChars, AttributeList attribs);
  void insert(String text, AttributeList attribs);
  void remove(int noChars);
\end{lstlisting}
This class allows services to specify changes they want to perform
in a sequential way e.g. keep the first 10 characters untouched, 
remove the next 5, insert text ``Hello World'' and apply attribute
[``bold'',''true''] to it, keep the next 2 characters unchanged but apply
attribute [``bold'',''''] to them etc. The ChangesetBuilder class will  
produce a correctly encoded changeset which can be then transmitted 
to \march{}. 

The last best practice I would like to share is a simple and efficient
algorithm of converting a list of changes of type ``{\it apply attribute [$key_i$, $value_i$] from character $begin_i$ to character $end_i$}'' to a sequential
list of changes suitable for the ChangesetBuilder. I used this 
algorithm (with minor changes) for all 4 services, hence might be of interest
for future service developers.
\begin{compactitem}
\item we create a list of ``event'' triples having signature $(type,
  i, attr)$ where $type$ is either ``add'' - to add attribute ``attr''
  to the list of attributes applied to all following
  characters, or ``remove'' to remove attr from the list of
  attributes. Index $i$ specifies at which position in the sequence 
  a certain event should take place.
\item for each rule of type ``{\it apply attribute [$key_i$,
    $value_i$] from character $begin_i$ to character $end_i$}'' add
  event triples (``add'', $begin_i$, [$key_i$, $value_i$]) and
  (``remove'', $end_i+1$, [$key_i$, $value_i$])
\item sort the event list by the $i$ values
\item initialize an empty list of attributes called {\it currentAttrs}
  and set lastPos = 0
\item iterate through the sorted event list and let $(type, i, attr)$
  be the current event
  \begin{itemize}
  \item if $i>lastPos$, add sequential operation keep($i-lastPos,
    currentAttrs$) and set $lastPos = i$.
  \item if $type$=''add'', add attr to currentAttrs
  \item if $type$=''remove'', remove attr from currentAttrs
  \end{itemize}
\end{compactitem}
 
This algorithm can be generalized to handle events which delete or insert text as well. 

\subsection{Evaluation of Integration Costs}
\label{extensibility}
The aim of the presented architecture is to minimize integration costs and hence
in this section I want to evaluate what we gain by using it. 

\subsubsection{Costs for Integrating Custom Editors / Services} 
~

Both editors and services need to implement the following functionality
\begin{compactenum}
\item [{\bf RE1.}] Connect to the \march{} component,
\item [{\bf RE2.}] Implement document model and changeset synchronization mechanisms.
\end{compactenum}
Both RE1 and RE2 can be reused from other already integrated 
editors/services. If no such implementations exist for a certain programming 
language, my own experience shows that
one needs to invest about one day for implementing RE1 and about two weeks for RE2. 
Implementing RE2 requires mostly code porting skills (for about 2k lines of code) 
i.e. good understanding of particularities of programming languages but does not require deep 
understanding of the algorithms themselves. Additionally, unit tests help a lot
finding and fixing bugs. Currently, RE1 and RE2 are available for 
Java and JavaScript languages using jeasysync2 \footnote{\url{https://github.com/jucovschi/jeasysync2}} 
and easysync2 \footnote{\url{https://github.com/Pita/etherpad-lite/static/js/Changeset.js}} libraries.

Editors need to additionally implement styling and user interaction mechanisms 
based on attributes in the shared document model. Depending on the editing environment, 
this might take several more days.

Integration of services requires implementation of the interaction of the service
with the shared document and with the user. To integrate each service described in section \ref{impl},
I needed 50-100 lines of code. 

\subsubsection{Requirements for Integrating a Custom Interaction}~

To add new user interactions, one has to choose a unique URI to identify the interaction
and then use this URI from an editor (to trigger events) or from services by using it
inside an attribute. The biggest cost associated with adding/extending the set of interactions is
that of propagating it to all already integrated editors. Versioning and change management 
of interaction URIs is an open issue.

%%% Local Variables: 
%%% mode: latex
%%% TeX-master: "paper"
%%% End: 

%% file: scenarios.tex
\subsection{Implemented Semantic Services}
\label{scenario}
  My architecture can be seen as an enabler for user-editor-service 
  type of interactions and hence this is the part which needs most testing.
  Let us consider the service of
  semantic syntax highlighting. The user-editor-service interaction 
  consists in the service being able to change the color of text. 
  Testing a more complex service, requiring coloring parts of 
  text  does not make sense because the additional service complexity 
  is independent of the presented architecture.  
  Hence I chose services testing different aspects of user-editor-service
  interaction, namely:
  \begin{compactenum}
  \item [{\bf \stex{} semantic syntax highlighter}] colors \stex{}
    code based on its semantic meaning. This is a service
    which cannot be implemented using regular expressions --- the main
    tool for syntax highlighting in many editors. Hence I implemented
    it as a service and integrated it in editors via \march{}. Even though it only
    needs to highlight text, it has to do that more often then most other
    services hence it helps benchmarking the user-editor-service 
    interaction speed. 
  \item [{\bf Term Spotter}] is a NLP based service which tries to
    spot mathematical terms inside a document. The interaction with the 
    user is very similar to that of spell checking, namely, spotted 
    mathematical terms are underlined while the user is typing. The
    user can then choose to add semantic references to spotted terms. 
    This is an example of a service with heavier server side part
    and helps us test how service results are automatically integrated 
    (if possible) into newer document versions. 
  \item [{\bf TermRef Hider and Transclusion}] services are examples
    of advanced editing features, one hiding 
    parts of annotations from the user and other showing referenced text instead 
    of references. As both services showcase very important results
    of using proposed architecture I address them in more detail in the next section.
  \end{compactenum}

  \subsubsection{Support for Advanced Editing Features}
  \label{outsource}
  ~
  
  Inline annotated documents are very hard to author because they contain 
  additional implicit knowledge that 
  \begin{inparaenum}
  \item is redundant to the author as she already knows it and 
  \item hinders a clear reading experience.
  \end{inparaenum}
  Stand-off or parallel annotations solve these problems but they require the 
  use of special editing environments every time a small update needs to be
  performed. Failing to do so may invalidate existing annotations. Generally 
  MKM systems only support inline annotations on documents that are editable 
  by users.

  In Figures \ref{stex-example-1} and \ref{stex-example-2} compare a mathematical 
  document to its semantically annotated version. The difference between the
  readability of these documents is quite obvious even though all we did was 
  to annotate three terms (using \textbackslash{}termref macros) and do four transclusions
  (using \textbackslash{}STRlabel and \textbackslash{}STRcopy macros).
\begin{figure}
\begin{lstlisting}[language=sTeX]
The gravitational potential energy of a system of masses $m_1$ and $M_2$ 
at a distance $r$ using gravitational constant $G$ is 
\begin{equation}
  U = -G\frac{m_1M_2}{r}+K
\end{equation}
where K is the constant of integration. Choosing the convention that $K=0$ 
makes calculations simpler, albeit at the cost of making U negative.
\end{lstlisting}
  \caption{Conventional mathematical document}
  \label{stex-example-1}

\begin{lstlisting}[language=sTeX]
The \termref{cd=physics-energy, name=grav-potential}{gravitational potential energy} 
of a system of masses \STRlabel[m1]{$m_1$} \STRcopy{m1} and \STRlabel[m2]{$M_2$} 
\STRcopy{m2} at a distance \STRlabel[r]{$r$} \STRcopy{r} using 
\termref{cd=physics-constants, name=grav-constant}{gravitational constant} 
\STRlabel[G]{$G$} \STRcopy{G} is \STRlabel[U]{$U$} \STRcopy{U}
\begin{equation}
  \STRcopy{U} = -\STRcopy{G}\frac{\STRcopy{m1}\STRcopy{m2}}{\STRcopy{r}}+
                \STRcopy{K}
\end{equation}
where \STRcopy{K} is the \termref{cd=physics-constants, name=integration}{constant 
of integration}. Choosing the convention that \STRcopy{K}$=0$ makes calculations 
simpler, albeit at the cost of making \STRcopy{U} negative.
\end{lstlisting}
  \caption{Semantically annotated mathematical document}
  \label{stex-example-2}

\end{figure}

To make the text in figure \ref{stex-example-2} look as readable 
as the one in \ref{stex-example-1}, we need to support 2 features,
namely: inline folding and transclusion. Using inline folding, one
could collapse a whole \textbackslash{}termref to show only the text
in its second argument. The transclusion feature would then replace 
\textbackslash{}STRcopy references with the text in the 
\textbackslash{}STRlabel. 

The inline folding or transclusion features are not supported by
most editing environments used to author MKM formats like \stex{}, 
Mizar\cite{UrbanBan:pem06} or LF\cite{Pfenning91}. Adding these features directly in each of the authoring 
environments requires a lot of initial development effort and incurs high
maintenance costs when the editor evolves.

%%% Local Variables: 
%%% mode: latex
%%% TeX-master: "paper"
%%% End: 

%% file: conclusion.tex
\section{Conclusion \& Future Work}
\label{concl}
The extent to which editing environments could support the authoring
process of MKM documents is far from being reached. There are lots
useful authoring services which, if integrated in editing environments, would
make authoring of MKM documents easier to learn, more efficient and
less error-prone. 

Many MKM editing services are available only in certain editing
environments but even more services live solely in the wish-lists
of MKM authors. An important reason for it is that creating and
maintaining such integrations is very expensive. This paper suggests
that integration of editors with MKM authoring services can be done in an
efficient way. Namely, a service showing the type of LF symbols
needs to be integrated with the \march{} component once and then 
be used in a ever-growing list of editors like Eclipse, jEdit or even
web-editors. Conversely, once an editor (e.g. \TeX{}macs)
integrates with the \march{} component, it would be able to provide
the user with all the services already integrated with the \march{}
component. 

The implementation part of this paper allowed me to test my ideas
and I found out that integrating a service into all already integrated 
editors can take as little as 3-4 hours of work. In case that
no communication and synchronization libraries are available for a
certain language, one can implement them in about 2-3 weeks, 

While the focus of the current paper is to reduce the costs for
integrating $m$ services into $n$ editors, the suggested solution
also:
\begin{compactenum}
\item makes is possible for services which need longer processing
  times to run in the background without interrupting the user's authoring
  experience,
\item allows services to be implemented in any convenient programming
  language or framework and even be distributed on different hardware,
  and
\item extends the typical plain-text document model with attributes
  which provide a very convenient storage for layers of semantic 
  information inferred by services. 
\end{compactenum}

Future research directions include development and integration of new
MKM services into editors, extending the list of programming languages 
which can connect to the \march{} component, and integrating additional
editors into the proposed architecture.

% what I want to do with it
% where it becomes relevant

%%% Local Variables: 
%%% mode: latex
%%% TeX-master: "paper"
%%% End: 